# QAM Adaptive Measurements Feedback Quantum Receiver Performance

Tian Chen, Ke Li, Yuan Zuo and Bing Zhu

***Abstract*—We theoretically study the quantum receivers with adaptive measurements feedback for discriminating quadrature amplitude modulation (QAM) coherent states in terms of average symbol error rate. For rectangular 16-QAM signal set, with different stages of adaptive measurements, the effects of realistic imperfection parameters including the sub-unity quantum efficiency and the dark counts of on-off detectors, as well as the transmittance of beam splitters and the mode mismatch factor between the signal and local oscillating fields on the symbol error rate are separately investigated through Monte Carlo simulations. Using photon-number-resolving detectors (PNRD) instead of on-off detectors, all the effects on the symbol error rate due to the above four imperfections can be suppressed in a certain degree. The finite resolution and PNR capability of PNRDs are also considered. We find that for currently available technology, the receiver shows a reasonable gain from the standard quantum limit (SQL) with moderate stages.**

## I. Introduction

Quantum mechanics sets fundamental limits on the attainable performance of the coherent state discrimination receiver. Nonorthogonal coherent states cannot be distinguished with total certainty because of their intrinsic overlap [1]. Efficient measurement and discrimination stratgies for coherent states are essential in many quantum information processing tasks, including communication, sensing and metrology, and various cryptographic protocols [2], [3], [4]. For conventional receivers (direct detection, homodyne and heterodyne receivers) [18], the discrimination error is bounded by the shot noise limit, which is often referred to as the standard quantum limit (SQL). On the other hand, Helstrom developed quantum detection theory and obtained the ultimate limit (Helstrom limit) on the average symbol error rate of coherent state discrimination in 1960s [1]. Though Helstrom provided a theory to find the ultimate limit of error probability, this mathematical specification does not usually translate into an explicit receiver specification realizable using standard optical components. And thus, many efforts have been devoted to physical implementation of quantum receivers to approach the Helstrom limit for different coherent state modulation sets.

For binary signals, the Kennedy receiver [5], the Dolinar receiver [6], the Optimal Displacement receiver [8] and the Partitioned-Interval Detection receiver [9], [10] have been investigated sequentially. Furthermore, most of these receivers for binary coherent state discrimination are experimentally demonstrated [7], [8]. For multiple modulation signals, quantum receivers based on adaptive measurements feedback or feedforward have also been proposed for M-ary phase shift keying (PSK) [11], [12], [13] and M-ary pulse position modulation (PPM) [14], [15], [16] signals.Another receiver scheme for M-ary PSK is a hybrid structure consisting of a homodyne receiver and a subsequent optimized displacement receiver using feed forward [17]. In practice, the sub-unity quantum efficiency and the dark counts of single-photon detectors, as well as the transmittance and the mode mismatch of beam splitters will deteriorate the error performances. Recently, the robustness against the dark counts [18] and the mode mismatch [19], with photon-number-resolving detectors (PNRD) instead of ON-OFF detectors was proved for M-ary PSK through both simulations and experiments [21]. These achievements have been noticed by fiber optic communication researchers. In fiber optic communication, quadrature amplitude modulation (QAM) signals are promising for further increasing the spectral efficiency [25]. Although in principle adaptive measurements quantum receivers can be used to discriminate arbitrary coherent states near the Helstrom limit, for QAM signals, there is no concrete and comprehensive receiver error performance analysis.

In this paper, for rectangular 16-QAM modulation signal set, the symbol error rates of adaptive measurements feedback quantum receiver were simulated for ON-OFF detectors and PNRDs. With different stages of adaptive measurement, the effects of above four imperfect devices' factors on the receiver performance were investigated. And it will be proved that, with PNRD instead of ON-OFF detectors, all the effects errors due to non-ideal devices can be suppressed effectively, especially in the regime of high signal mean photon numbers and small number of feedback stages. Besides, the effect of the finite resolution and PNR capability are also considered here.

## II. *M*-ary QAM System and Its Performance Limits

In this section, let us review the quantum state characterization, the SQL, the Helstrom limit [23] and square root measurement (SRM) [22], [24] of *M*-ary QAM signals.

Ke Li is with the Nanjing Research Institute of Electronic Technology, Nanjing, Jiangsu 210000, China (e-mail: likeniao@mail.ustc.edu.cn).

The other authors are with the Department of Electronic Engineering and Information Science, University of Science and Technology of China, Hefei, Anhui 230027, China (e-mail: chentian@mail.ustc.edu.cn; zuoyuan@mail.ustc.edu.cn; (corresponding author) zbing@ustc.edu.cn).

### *A. Quantum State Characterization of M-ary QAM Signals*

In quantum theory, if the annihilation and the creation operators are $\hat{a}$ and $\hat{a}^\dagger$, QAM signals can be characterized by two quadrature amplitudes $\hat{x}_c$ and $\hat{x}_s$, which are defined as follows:

$$\hat{x}_c \equiv (\hat{a}+\hat{a}^\dagger)/2, \qquad \hat{x}_s \equiv (\hat{a}-\hat{a}^\dagger)/2j \tag{1}$$

where $j=\sqrt{-1}$ . For rectangular QAM, each quadrature amplitudes $\hat{x}_c$ and $\hat{x}_s$ can take $L$ values independently. Thus the number of signals $M$ is represented by

$$M = L^2, \qquad L = 3,4,5,\cdots. \tag{2}$$

For convenience, we define the index set $\Omega$ as follows:

$$\Omega = \{-(L-1)+2(i-1) \mid i = 1,2,\cdots,L\}. \tag{3}$$

Then the rectangular QAM signals can be defined as a set of coherent states:

$$\left|\varphi_{p,q}\right\rangle = \left|\alpha(p+jq)\right\rangle, \qquad p,q\in\Omega \tag{4}$$

Without loss of generality, $\alpha$ can be taken as a real number. Thus the average photon number $N_s$ is defined as

$$N_s = \sum_{p,q} P_{p,q}\left|\alpha(p+jq)\right|^2 \tag{5}$$

where $P_{p,q}$ is *a priori* probabilities of the QAM signals $\left|\varphi_{p,q}\right\rangle$.

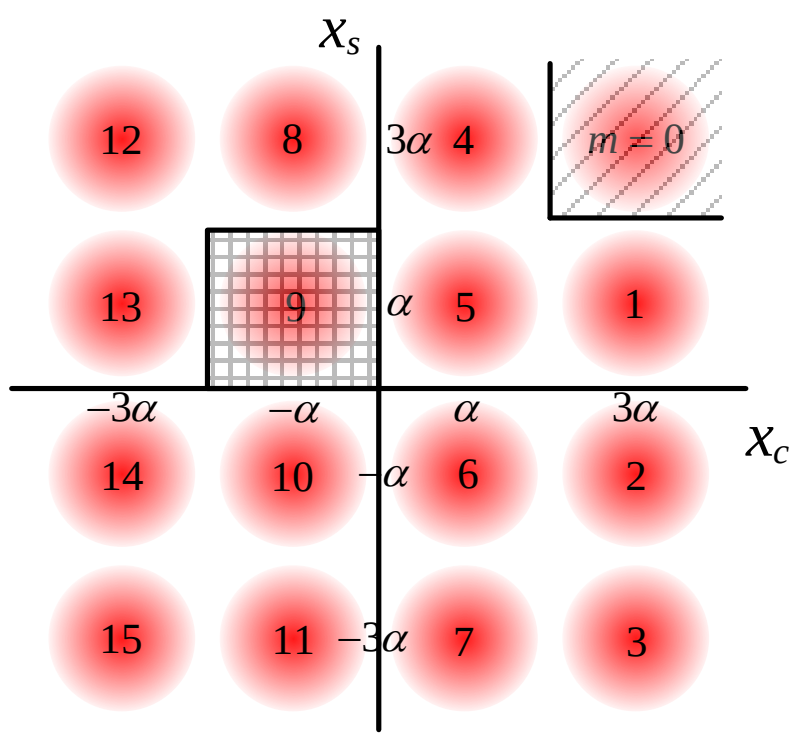

Fig. 1. Constellations of the rectangular 16-QAM

As a concrete example, we examine rectangular 16-QAM. In this case, the index set is $\Omega = \{-3,-1,1,3\}$. Then signals are denoted as follows:

$$\begin{aligned}
&\left|\varphi_{3,3}\right\rangle = \left|\alpha(3+3j)\right\rangle, && \left|\varphi_{3,1}\right\rangle = \left|\alpha(3+j)\right\rangle,\\
&\left|\varphi_{3,-1}\right\rangle = \left|\alpha(3-j)\right\rangle, && \left|\varphi_{3,-3}\right\rangle = \left|\alpha(3-3j)\right\rangle,\\
&\left|\varphi_{1,3}\right\rangle = \left|\alpha(1+3j)\right\rangle, && \left|\varphi_{1,1}\right\rangle = \left|\alpha(1+j)\right\rangle,\\
&\left|\varphi_{1,-1}\right\rangle = \left|\alpha(1-j)\right\rangle, && \left|\varphi_{1,-3}\right\rangle = \left|\alpha(1-3j)\right\rangle,\\
&\left|\varphi_{-1,3}\right\rangle = \left|\alpha(-1+3j)\right\rangle, && \left|\varphi_{-1,1}\right\rangle = \left|\alpha(-1+j)\right\rangle,\\
&\left|\varphi_{-1,-1}\right\rangle = \left|\alpha(-1-j)\right\rangle, && \left|\varphi_{-1,-3}\right\rangle = \left|\alpha(-1-3j)\right\rangle,\\
&\left|\varphi_{-3,3}\right\rangle = \left|\alpha(-3+3j)\right\rangle, && \left|\varphi_{-3,1}\right\rangle = \left|\alpha(-3+j)\right\rangle,\\
&\left|\varphi_{-3,-1}\right\rangle = \left|\alpha(-3-j)\right\rangle, && \left|\varphi_{-3,-3}\right\rangle = \left|\alpha(-3-3j)\right\rangle.
\end{aligned} \tag{6}$$

In phase space, the rectangular 16-QAM constellations are given in Fig. 1. For general rectangular M-ary QAM, if the same number criterion is used as in Fig. 1, QAM signals $\left|\varphi_{pq}\right\rangle$ with a priori probabilities $P_{p,q}$ can be denoted as $\left|\varphi_m\right\rangle$ with *a priori* probabilities $P_m$, where $m = 0,1,2,\cdots,M-1$.

### *B. Standard Quantum Limit of M-ary QAM Signals*

It is well known that the SQL of *M*-ary QAM signals can be obtained by using the heterodyne receiver. The heterodyne receiver simultaneously measures the two quadrature amplitudes $\hat{x}_c$ and $\hat{x}_s$. Assuming there is no thermal noise, the probability density function (pdf) of the heterodyne receiver is given as follows:

$$\begin{aligned}
p(x_{c,}x_s \mid p,q) &= \frac{1}{\pi}\operatorname{tr}\left(\left|\varphi_{p,q}\right\rangle\left\langle\varphi_{p,q}\right|\left|\alpha(x_{c,}+jx_s)\right\rangle\left\langle\alpha(x_{c,}+jx_s)\right|\right)\\
&= \frac{1}{\pi}\left|\left\langle\alpha(x_{c,}+jx_s)\middle|\alpha(p+jq)\right\rangle\right|^2\\
&= \frac{1}{\pi}\exp[-(x_c-p\alpha)^2-(x_s-q\alpha)^2]
\end{aligned} \tag{7}$$

where $p,q\in\Omega$.

According to classical detection theory, the conditional probability of detecting $\left|\varphi_{p',q'}\right\rangle$ when $\left|\varphi_{p,q}\right\rangle$ is prepared is:

$$P(p',q' \mid p,q) = \iint_{D_{p',q'}} p(x_{c,}x_s \mid p,q)\,dx_c dx_s \tag{8}$$

where $D_{p',q'}$ is the detection domain, and can be represented as:

$$D_{p',q'} = \{(x_{c,}x_s) \mid D_L(p') < x_c \le D_U(p'), D_L(q') < x_s \le D_U(q')\}$$

$$\begin{aligned}
D_L(p') &\equiv \begin{cases} -\infty, & p' < -L+2 \\ \alpha(p'-1), & p' > -L+2 \end{cases}\\
D_U(p') &\equiv \begin{cases} \infty, & p' > L-2 \\ \alpha(p'+1), & p' < L-2 \end{cases}
\end{aligned} \tag{9}$$

As illustrated in Fig. 1, the detection domain $D_{3,3}$ and $D_{-1,1}$ are the stripe region and the grid region, respectively. Thus the SQL for the symbol error rate of *M*-ary QAM signals is

$$P_{e_SQL} = 1-\sum_{p,q} P_{p,q} P(p,q \mid p,q) \tag{10}$$

### *C. Helstrom Limit of M-ary QAM Signals*

For *M*-ary QAM signals, it is difficult to obtain the Helstrom limit analytically. But according to the necessary and sufficient conditions for the optimum positive operator-valued measures (POVM) minimizing the symbol error rate and its dual formulation, the Helstrom limit can be obtained by using semidefinite programming.

The quantum states of *M*-ary QAM signals can be represented by a set of *M* density operators

$$\{\rho_m = \left|\varphi_m\right\rangle\left\langle\varphi_m\right|, 0 \le m \le M-1\} \tag{11}$$

which are positive semidefinite (PSD) and Hermitian on an *n*-dimensional complex Hilbert space *H*. In terms of the orthonormal basis of the number eigenstates $\left|n\right\rangle$, the coherent states $\left|\varphi_m\right\rangle$ take the forms

$$\left|\varphi_m\right\rangle = e^{-\frac{1}{2}\left|\varphi_m\right|^2}\sum_{n=0}^{+\infty}\frac{(\varphi_m)^n}{(n!)^{1/2}}\left|n\right\rangle \tag{12}$$

and $\left\langle\varphi_m\right|$ take the forms

$$\langle\varphi_m| = e^{-\frac{1}{2}|\varphi_m|^2}\sum_{n=0}^{+\infty}\frac{(\varphi_m^*)^n}{(n!)^{1/2}}\langle n| \tag{13}$$

Thus the infinite matrix representation of the density operator $\|\rho_m\|$ cna be obtained, and the expression of the $i\,j$ entry is

$$\rho_{m_i,j} = e^{-|\varphi_m|^2}\frac{(\varphi_m)^i}{(i!)^{1/2}}\frac{(\varphi_m^*)^j}{(j!)^{1/2}} \tag{14}$$

At the receiver, a measurement comprising *m* PSD Hermitian measurement operators $\{\Pi_m, 0\le m\le M-1\}$ is constructed on *H*. With *a priori* probabilities $P_m$ for each $\rho_m$, the probability of correct detection is given by

$$P_d = \sum_{m=0}^{M-1} P_m\,\mathrm{Tr}(\rho_m\Pi_m) \tag{15}$$

Thus, form quantum detection theory, the Helstrom limit can be obtained by solving the following maximization problem:

$$\max_{\Pi_m\in B}\sum_{m=0}^{M-1}\mathrm{Tr}(\rho_m'\Pi_m) \tag{16}$$

subject to the constraints

$$\begin{gathered}\Pi_m \ge 0,\quad 0\le m\le M-1\\ \sum_{m=0}^{M-1}\Pi_m = I.\end{gathered} \tag{17}$$

where *B* is the set of Hermitian operators on *H* and $\rho_m' = P_m\rho_m$. After obtaining the optimal POVM $\{\hat{\Pi}_m, 0\le m\le M-1\}$, the Helstrom limit of the symbol error rate can be obtained as:

$$P_{e_Helstrom} = 1-\hat{P}_d,\quad \hat{P}_d = \sum_{m=0}^{M-1}P_m\,\mathrm{Tr}(\rho_m\hat{\Pi}_m) \tag{18}$$

The dual problem of the above maximization problem has the following formulation:

$$\min_{X\in B}\ \mathrm{Tr}(X) \tag{19}$$

subject to

$$X \ge \rho_m',\quad 0\le m\le M-1 \tag{20}$$

We should note that this problem is a Complex-Valued Linear Matrix Inequalities optimization problem. By using MATLAB LMI Control Toolbox, the optimal solution $\hat{X}$ can be obtained.

Then the Helstrom limit of the symbol error rate is

$$P_{e_Helstrom} = 1-\mathrm{Tr}(\hat{X}) \tag{21}$$

It should be noted that the density matrix $\|\rho_m\|$ given in (14) is infinite dimensional. But we need a finite dimensional density matrix when solving the optimation problem by using MATLAB. So the density matrix given in (14) need to be truncated. Considering a density operator has a unitary trace, we can use the following criterion to truncate $\|\rho_m\|$. We choose the smallest integer $N_\varepsilon$ such that

$$\sum_{i=0}^{N_\varepsilon-1}\rho_{m_i,i} > 1-\varepsilon \tag{22}$$

where $\varepsilon$ is the required accuracy.

With increasing the average photon number $N_s$, the truncation integer $N_\varepsilon$ becomes more and more larger. Thus solving the optimization problem becomes more and more time and memory consuming.

### *D. Square Root Measurement of M-ary QAM Signals*

From the above analysis, we know that in quantum detection theory if we want to obtain the Helstrom limit, we should search the optimum POVM minimizing the symbol error rate.

Unlike the semidefinite programming, the optimum or the asymptotical optimum POVM can be obtained through the SRM systematically and efficiently from the prepared signal set $\{\rho_m = |\varphi_m\rangle\langle\varphi_m|, 0\le m\le M-1\}$. For quantum states with geometrically uniform symmetry, such as M-ary PSK signals and M-ary PPM signals, it gives the optimum POVM. While for QAM signals, it gives asymptotical optimum POVM, which means it becomes almost optimum when the average photon number of signals is increased. In practice, we can use the results obtained from SRM to speculate the position and trends of Helstrom limit efficiently.

The SRM is defined as follows:

$$\begin{gathered}\hat{\Pi}_m = |\mu_m\rangle\langle\mu_m|\\ |\mu_m\rangle = \hat{G}^{-1/2}|\varphi_m\rangle\\ \hat{G} = \sum_{m=0}^{M-1}|\varphi_m\rangle\langle\varphi_m|\end{gathered} \tag{23}$$

where $\hat{G}$ is the Gram operator. The conditional probability adopting the SRM for detecting signal *j* when signal *i* was sent is

$$P(j\,|\,i) = |\langle\mu_j|\varphi_i\rangle|^2 = |\langle\mu_j|\hat{G}^{1/2}|\varphi_i\rangle|^2 \tag{24}$$

By using operator algebra, it is difficult to calculate the square root of the Gram operator $\hat{G}^{1/2}$. However, the problem can be solved by the matrix analysis. The following gives the detailed algorithm.

First, we define the Gram matrix $G$ which is a matrix representation of the operator $\hat{G}$ as follows:

$$G = \begin{bmatrix}\langle\varphi_0|\varphi_0\rangle & \cdots & \langle\varphi_0|\varphi_{M-1}\rangle\\ \vdots & \ddots & \vdots\\ \langle\varphi_{M-1}|\varphi_0\rangle & \cdots & \langle\varphi_{M-1}|\varphi_{M-1}\rangle\end{bmatrix} \tag{25}$$

Each element in the Gram matrix is the inner product of two coherent states $\langle\alpha|\beta\rangle = \exp\left[\alpha^*\beta - |\alpha|^2/2 - |\beta|^2/2\right]$. For QAM signals, the inner products are represented by

$$\langle\varphi_{p',q'}|\varphi_{p,q}\rangle = \exp\left[-\frac{\alpha^2}{2}\left\{(p'-p)^2+(q'-q)^2\right\} + j\alpha^2(p'q-q'p)\right] \tag{26}$$

According to the number criterion in Fig. 1, the Gram matrix $G$ for QAM signals can be obtained.

Then we calculate the eigenvalues $\lambda_i$ and the corresponding normalized eigenvectors $\boldsymbol{\lambda}_i$ of the Gram matrix $G$. By using these eigenvalues and eigenvectors, the next matrices can be defined

$$\begin{gathered}D^{1/2} = \mathrm{diag}\left[\sqrt{\lambda_0}, \sqrt{\lambda_1}, \cdots, \sqrt{\lambda_{M-1}}\right]\\ Q = \left[\boldsymbol{\lambda}_0, \boldsymbol{\lambda}_1, \cdots, \boldsymbol{\lambda}_{M-1}\right]\end{gathered} \tag{27}$$

Then the square root of the Gram matrix is $\hat{G}^{1/2}$ obtained by

$$G^{1/2} = QD^{1/2}Q^{?} \tag{28}$$

And the conditional probability can be calculated as

$$P(j\,|\,i) = \left|(G^{1/2})_{ji}\right|^2 \tag{29}$$

Thus the symbol error rate adopting SRM is

$$P_{e_SRM} = 1 - \sum_{m=0}^{M-1} P(m\,|\,m) \tag{30}$$

After reviewing the performance limits for QAM signals in this section, we then focus on the physical implementation of QAM quantum receiver in the following.

## III. Adaptive Measurements Quantum Receiver

As we have already known, arbitrary coherent states set can be discriminated near the Helstrom limit by using adaptive measurements quantum receivers. But for QAM signals, there is no concrete and comprehensive analysis. In this section, we will fully study the performances of the QAM adaptive measurements feedback quantum receiver through Monte Carlo simulations.

### *A. Receiver Configuration and Modeling*

Let us first introduce the configuration of the adaptive measurements quantum receiver. Fig. 2 illustrates the adaptive measurements quantum receiver in the feedback form. It should be noted that adaptive measurements can also be implemented in the feedforward form, which is not showed here.

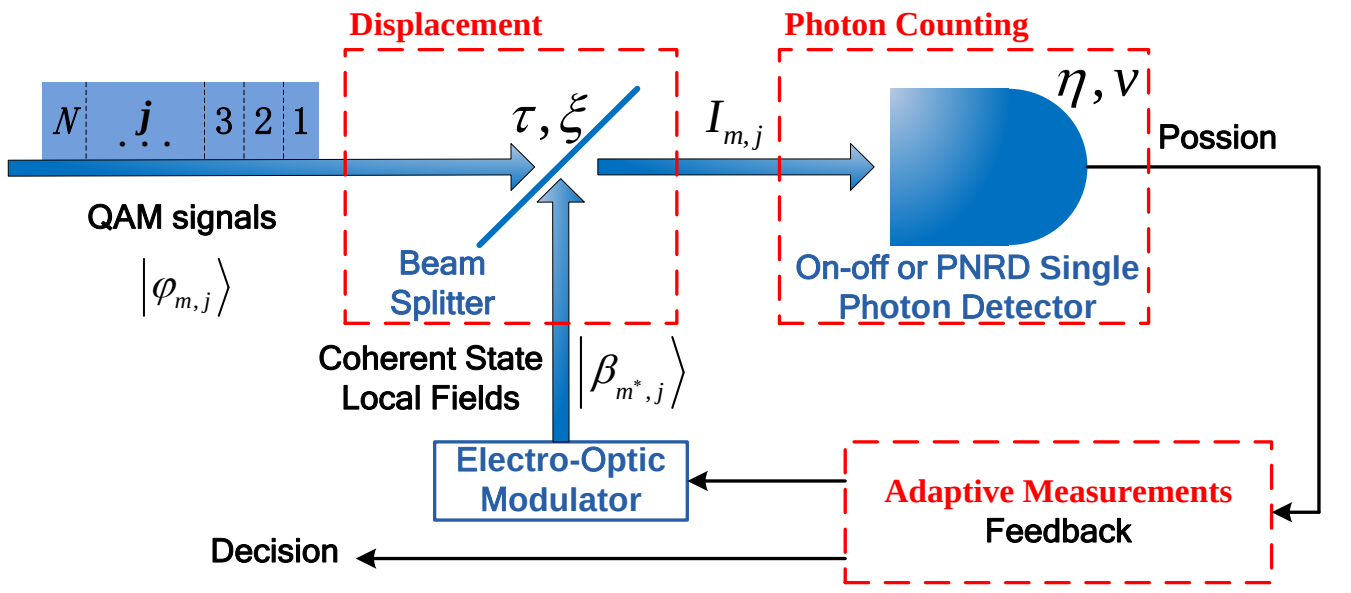

Fig. 2. The configuration of the adaptive measurements feedback quantum receiver.

As illustrated in Fig. 2, the adaptive measurements quantum receiver is constructed by three important parts: displacement operation, photon counting and adaptive measurements.

Throughout this paper, for each received symbol $|\varphi_m\rangle$ ,we assume *a priori* probabilities of the signals are all the same, i.e., $P_m = 1/M$ .

Each received signal codeword interval is partitioned into $N$ consecutive disjoint equal durations with indexes $j = 1, 2, 3, \cdots, N$ , and the signals corresponding to each duration are $|\varphi_{m,j}\rangle = |\varphi_m/\sqrt{N}\rangle$ .After the partition, the signal in the *j*th duration $|\varphi_{m,j}\rangle$ is displaced by a local field $|\beta_{m^*,j}\rangle$ via a beam splitter. According to the model [20], assuming that there is only one mode in both $|\varphi_{m,j}\rangle$ and $|\beta_{m^*,j}\rangle$ , the average intensity $I_{m,j}$ of the field after the beam splitter in the photon number units is

$$I_{m,j} = (1-\xi)\tau\left|\varphi_{m,j}\right|^2 + \xi\left|\sqrt{\tau}\varphi_{m,j} - \beta_{m^*,j}\right|^2 \tag{31}$$

where $\tau$ describes the transmittance of the beam splitter and $\xi$ describes the mode mismatch due to the imperfect interference between $|\varphi_{m,j}\rangle$ and $|\beta_{m^*,j}\rangle$ . In the equation (31), $|\beta_{m^*,j}\rangle$ satisfies

$$\beta_{m^*,j} = \begin{cases} \sqrt{\tau}\varphi_{0,1} = \sqrt{\tau}\varphi_0/\sqrt{N} \\ \sqrt{\tau}\varphi_{m^*,j-1} \end{cases} \tag{32}$$

where $m^*$ is the temporary optimum decision, which is given by the adaptive measurement strategy.

After the displacement operation, the displaced fields is detected by photon counting. And the probabilities corresponding to the photon counting events detecting $n_j$ photons in the jth duration can be denoted as $P_{m,j}(n_j)$ . Here the single-photon detector can be ON-OFF detectors or PNRDs. If ON-OFF detectors are used, $P_{m,j}(n_j)$ is

$$P_{m,j}(n_j) = \begin{cases} e^{-\nu-\eta I_{m,j}}, n_j = 0 \\ 1 - e^{-\nu-\eta I_{m,j}}, n_j \neq 0 \end{cases} \tag{33}$$

where $\nu$ describes the dark counts and $\eta$ describes the quantum efficiency of the single-photon detector. If PNRDs with finite photon-number-resolution $n_{PNR}$ are used, $P_{m,j}(n_j)$ becomes

$$P_{m,j}(n_j) = \begin{cases} e^{-\nu-\eta I_{m,j}} \dfrac{(\nu+\eta I_{m,j})^{n_j}}{n_j!}, n_j = 0, 1, 2, \cdots, n_{PNR} \\ 1 - \sum\limits_{n_j=0}^{n_{PNR}} e^{-\nu-\eta I_{m,j}} \dfrac{(\nu+\eta I_{m,j})^{n_j}}{n_j!}, n_j > n_{PNR} \end{cases} \tag{34}$$

From equations (33) and (34), we know that ON-OFF detectors are equivalent to PNRD with $n_{PNR} = 0$ . And for PNRD ignoring the effects of finite photon-number-resolution, which means $n_{PNR} = +\infty$ , $P_{m,j}(n_j)$ becomes

$$P_{m,j}(n_j) = e^{-\nu-\eta I_{m,j}} \frac{(\nu+\eta I_{m,j})^{n_j}}{n_j!}, n_j = 0, 1, 2, \cdots, +\infty \tag{35}$$

Finally, after photon counting, adaptive measurements are implemented to update the local field $|\beta_{m^*,j+1}\rangle$ in the (*j*+1)th duration. Here the maximum *a posteriori* (MAP) criterion is used as the adaptive measurement strategy. After detecting $n_j$ photons in the *j*th duration, the *a posteriori* probabilities are given by

$$P_{post_m,j} = \frac{P_{prior_m,j} \cdot P_{m,j}}{\sum_{l=0}^{M-1} P_{prior_l,j} \cdot P_{l,j}} \tag{36}$$

The *a priori* probability $P_{prior_m,j}$ in the *j*th duration satisfies

$$P_{prior_m,j} = \begin{cases} P_m, j = 1 \\ P_{post_m,j-1}, j > 1 \end{cases} \tag{37}$$

Then $m^*$ in the local field $|\beta_{m^*,j+1}\rangle$ can be obtained by

$$P_{post_m^*,j} = \max_m \left(P_{prior_m,j}\right) \tag{38}$$

And $m^*$ in the last duration corresponding to $P_{post_m^*,N}$ is the decision output.

### B. *Receiver Performances Simulations*

When the number of the durations $N$ is large and the devices' imperfections are considered, the adaptive measurements strategy used here makes it complicated to obtain the receiver symbol error rate analytically. Therefore, after the above probability model is established, Monte Carlo simulations are used to analyze the receiver performances. With different stages of adaptive measurements, i.e. the number of partitions N is different, the separate effects of the four imperfect parameters including the sub-unity quantum efficiency, the dark counts of photon counting, the transmittance of beam splitters and the mode mismatch factor on the receiver error performance are investigated as shown in Fig. 3 to Fig. 6. The effect of the finite resolution and PNR capability are studied as shown in Fig. 7. Each plot is given by a Monte Carlo simulation with $10^6$ trials. In each figure, the black solid and dashed lines represent the SRM bound and the standard quantum limit, respectively.

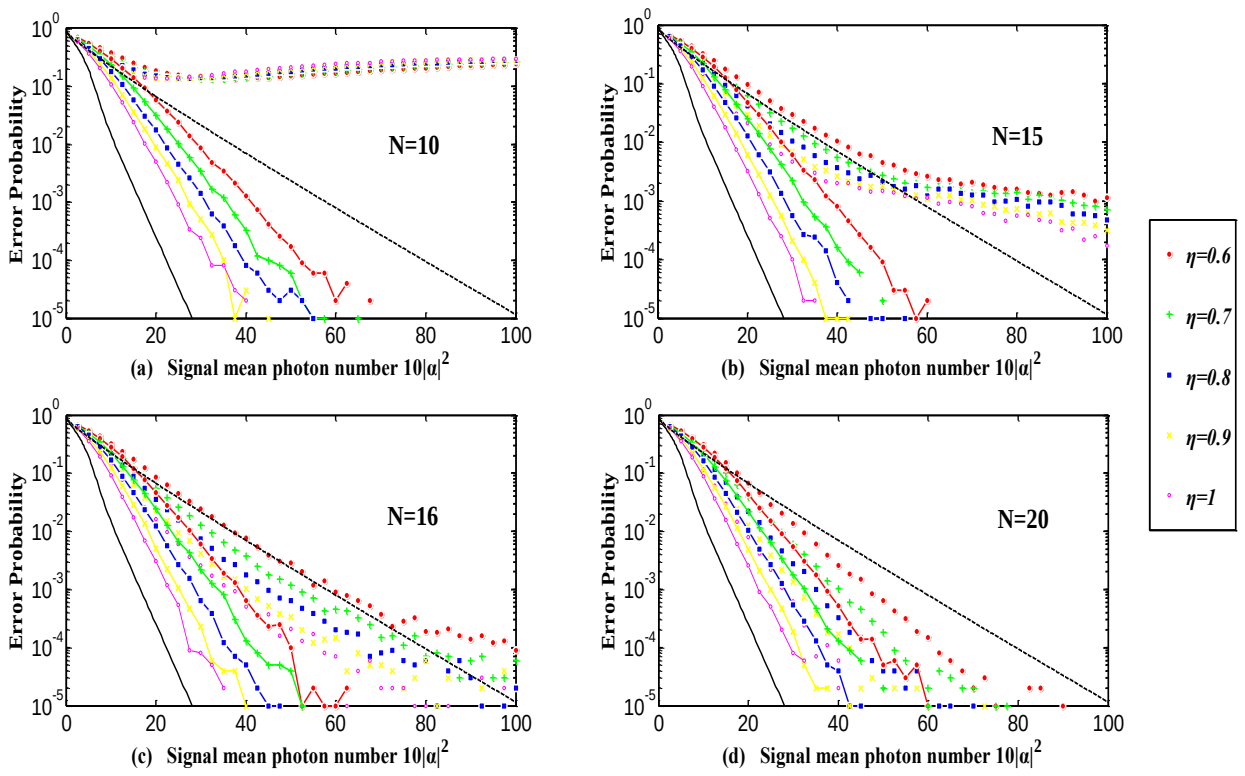

Fig. 3. The separate effects of the quantum efficiency ( $\eta$ varies with $v=0$ , $\tau=1$ , and $\xi=1$ ) of the single-photon detector with the different number of partitions (a) N=10 and (b) N=15, as well as (c) N=16 and (d) N=20 for rectangle 16-QAM quantum receiver.

Fig. 3 to Fig. 6 illustrate the simulation results for rectangle 16-QAM signals. The receiver performances are shown with the ON-OFF detection (colored dotted) and the PNRD ( $n_{PNR}=+\infty$ ) detection (colored solid). Each shows the separate effects of the four imperfect factors on the receiver error probability, respectively. All these pictures show that receiver error rate is reduced as the number of partitions increases. When $N<M$ , the receiver performances benefit greatly from the increase of the number of partitions for on-off detectors whereas the receiver performances for PNRDs gain much smaller compared to which for on-off detectors. When $N>M$ , the error probabilities for both detectors change smoothly despite of the variation of $N$ .

From Fig. 3 and Fig. 5, it can be derived that the quantum efficiency of the single-photon detector and the transmittance of the beam splitter have the same effect on the receiver performance, which is in coincidence with the formulae of the established probability model. From Fig. 4, we note that the error rates are saturated for on-off detectors as the signal mean photon numbers increase, which implies that the dark counts limit the performance of the receiver. From Fig. 6, it is shown that the receiver performance deteriorates seriously even with a minor mode mismatch for on-off detectors. So in practice, we should give more attention to the mode mismatch between the signal and the local oscillating fields. However, the four imperfect factors can be suppressed remarkably by using PNRDs instead of on-off detectors, especially for smaller number of partitions and higher signal mean photon numbers.

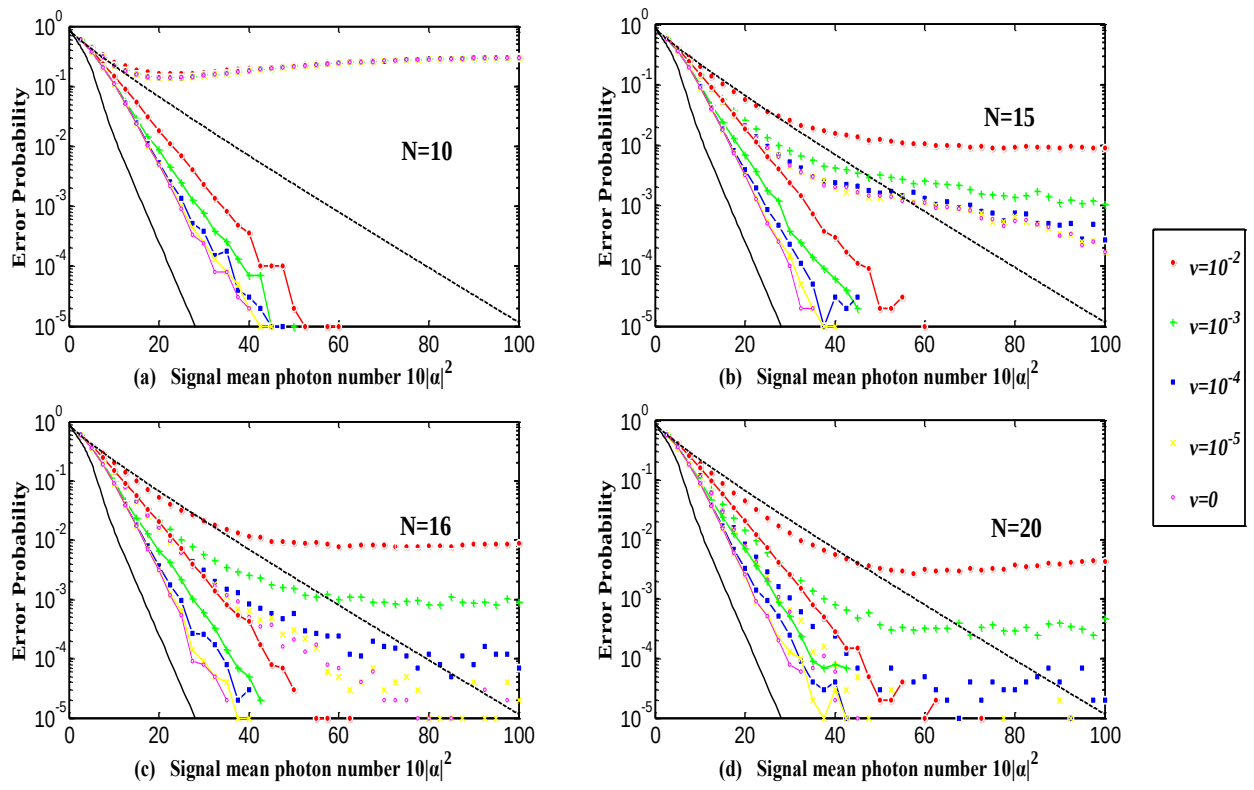

Fig. 4. The separate effects of the dark counts ( $v$ varies with $\eta=1$ , $\tau=1$ , and $\xi=1$ ) of the single-photon detector with the different number of partitions (a) N=10 and (b) N=15, as well as (c) N=16 and (d) N=20 for rectangle 16-QAM quantum receiver.

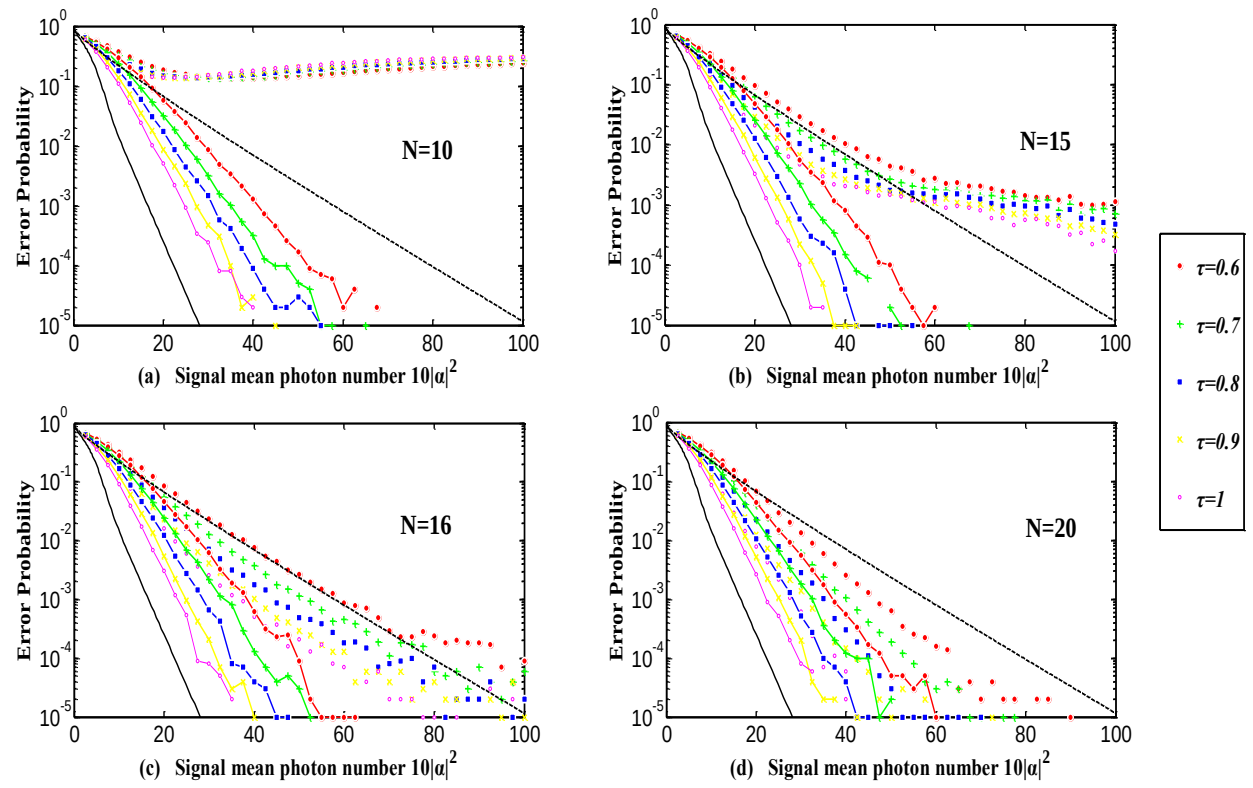

Fig. 5. The separate effects of the transmittance ( $\tau$ varies with $\eta=1$ , $v=0$ , and $\xi=1$ ) of the beam splitter with the different number of partitions (a) N=10 and (b) N=15, as well as (c) N=16 and (d) N=20 for rectangle 16-QAM quantum receiver.

Fig. 7 shows the effect of PNR capability on the performance of adaptive measures feedback quantum receiver with realistic imperfect factors [9] for 16-QAM. Colored solid lines represent the receiver performances of the PNRD detection with different $n_{PNR}$ . It is clear that the error probability decreases as the photon number resolution increases.We note that the error rates for PNRD-based recriver show steplike curves, due to discrete nature of photon number, such a classification varies discretely as a function of $|\alpha|^2$ . Apparently, it is impossible by on-off detectors ( $n_{PNR}=0$ ) that have the same number of feedback steps outperforming the SQL as that by PNRD with a moderate PNR capability.

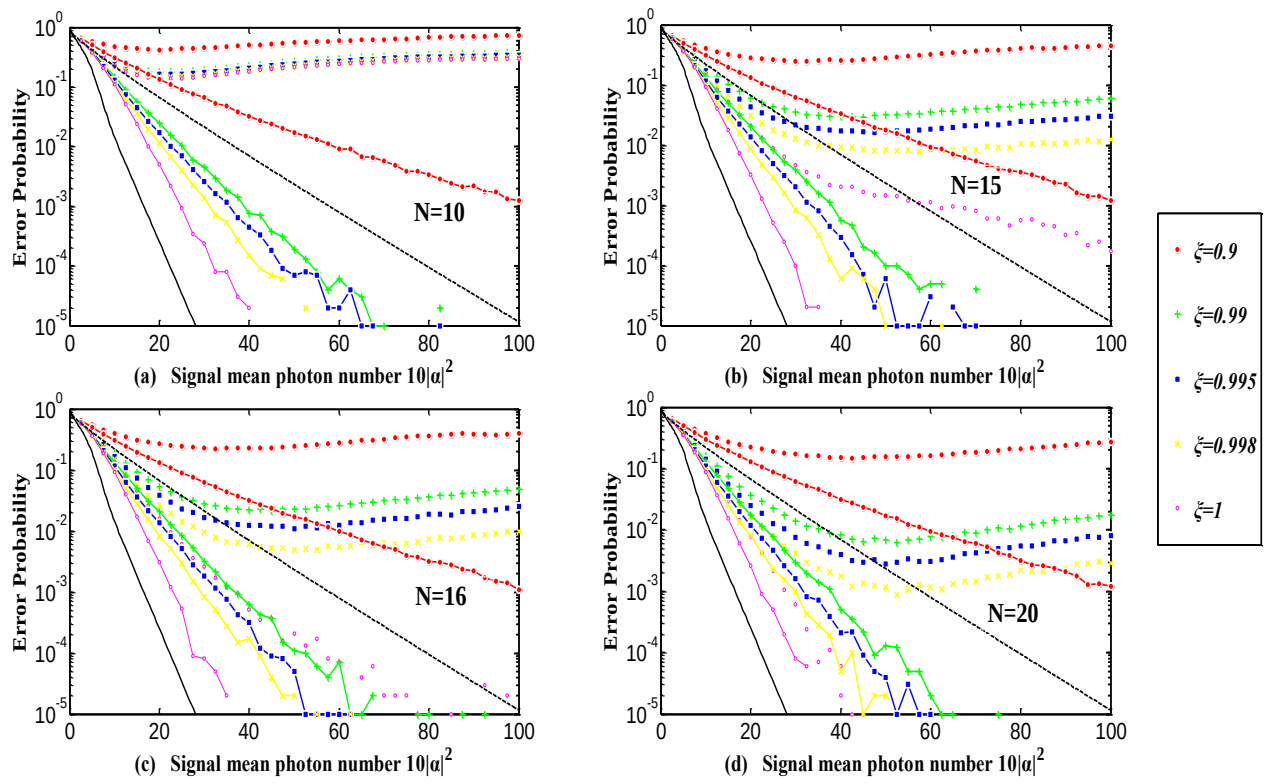

Fig. 6. The separate effects of the mode mismatch factor ( $\xi$ varies with $\eta = 1$, $v = 0$, and $\tau = 1$ ) of the beam splitter with the different number of partitions (a) N=10 and (b) N=15, as well as (c) N=16 and (d) N=20 for rectangle 16-QAM quantum receiver.

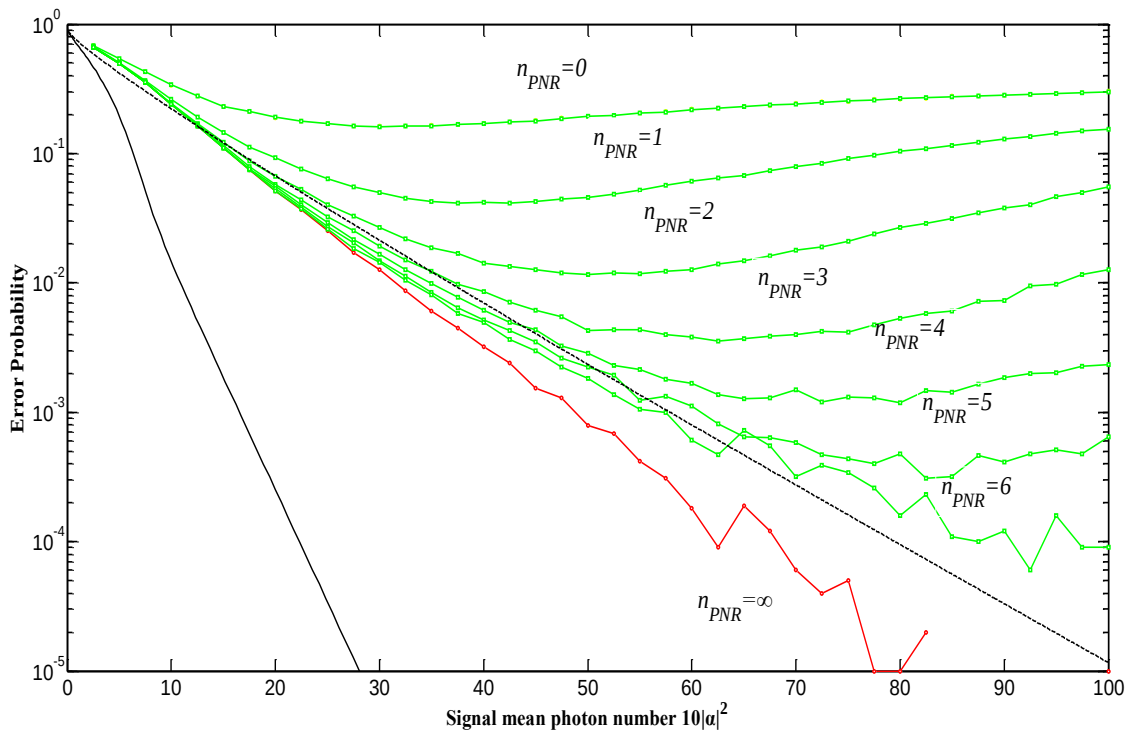

Fig. 7. The effect of PNR capability of PNRDs with the same number of partitions (N=10) for rectangle 16-QAM quantum receiver ( $\eta = 0.723$, $v = 2.7\times10^{-5}$, $\tau = 0.99$, and $\xi = 0.995$ ).

## IV. Conclusions and Discussions

In this letter, it is shown that the error rates of adaptive measures feedback quantum receiver can outperform the SQL for QAM. But the number of partitions must be large enough (at least $N = M$ ) if better performances of on-off detectors are expected which limits the signal repetition rate. When the modulation signal number is large, the practical application of the on-off-type adaptive measures feedback quantum receiver becomes quite complex to satisfy the need of the electrical bandwidth. In this situation, PNRD-based receiver with lower PNR capability will be a smart choice owing to its robustness against realistic non-ideal devices.